\documentclass[12pt]{article}

\usepackage{indentfirst}
\usepackage{amsmath}
\usepackage{amssymb}
\usepackage{amsfonts}
\usepackage{amscd}
\usepackage{amsbsy}
\usepackage{amsthm}
\usepackage{latexsym}
\usepackage{graphicx,color} 
\usepackage[dvipsnames]{xcolor}
\usepackage[colorlinks]{hyperref}
\hypersetup{linkcolor=blue,citecolor=blue,urlcolor=blue}
\def\nn{\nonumber}       
\def\beq{\begin{eqnarray}}
\def\eeq{\end{eqnarray}}
\def\ln{\,\mbox{ln}\,}

\def\diag{\,\mbox{diag}\,}
\def\Tr{\,\mbox{Tr}\,}

\DeclareMathOperator{\cx}{\square}

\def\al{\alpha}
\def\be{\beta}

\def\ga{\gamma}
\def\de{\delta}

\def\ep{\epsilon}
\def\ze{\zeta}

\def\la{\lambda}
\def\na{\nabla}
\def\pa{\partial}

\def\ph{\varphi}

\def\th{\theta}
\def\te{\vartheta}

\def\Ga{\Gamma}
\def\De{\Delta}

\usepackage[symbol]{footmisc} 
\usepackage{float} 
\usepackage{cite}  
\usepackage{titlesec}

\usepackage{ulem} 

\titleformat*{\section}{\large\bfseries}
\titleformat*{\subsection}{\normalsize\bfseries}

\begin{document}


\begin{center}
\renewcommand*{\thefootnote}{\fnsymbol{footnote}}
{\Large 
On the renormalization of massive vector field theory
\\
coupled to scalar in curved space-time}
\vskip 2mm

    {\bf Ioseph L. Buchbinder}$^{a,b}$
    \footnote{E-mail address: \ buchbinder@theor.jinr.ru},
\qquad
    {\bf P\'ublio Rwany B. R. do Vale}$^{c,d}$
    \footnote{E-mail address: \ publio.vale@gmail.com},
\\
    {\bf Guilherme Y. Oyadomari}$^{e,c}$
    \hspace{-1mm}\footnote{E-mail address: \  yoshioyadomari@hotmail.com},
\qquad
{\bf and} \
\qquad
{\bf Ilya L. Shapiro }$^{c,e}$
\footnote{E-mail address: \ ilyashapiro2003@ufjf.br}
\vskip 8mm

a) Bogoliubov Laboratory of Theoretical Physics, Joint
Institute of Nuclear Research,
141980, Dubna, Russia
\vskip 1mm

b) Tomsk State Pedagogical University,
637041, Tomsk, Russia
\vskip 1mm

c) Departamento de F\'{\i}sica, ICE, Universidade
Federal de Juiz de Fora,
\\
36036-900, Juiz de Fora, Minas Gerais, Brazil
\vskip 1mm

d)  Centro de Astropart\'iculas y F\'isica de Altas Energ\'ias,
Departamento de F\'isica Te\'orica,
Universidad de Zaragoza, E-50009 Zaragoza, Spain
\vskip 1mm

e)  PPGCosmo, Universidade Federal do Esp\'{\i}rito Santo,
Vit\'oria, 29075-910, ES, Brazil
\end{center}
\vskip 2mm


\begin{abstract}

\noindent
We consider the renormalization of massive vector field interacting
with charged scalar field in curved spacetime. Starting with the theory
minimally coupled to external gravity and using the formulations
with and without St\"{u}ckelberg fields, we show that the
longitudinal mode of vector field is completely decoupled and the
remaining theory of transverse vector field is renormalizable by
power counting.  The formal arguments based on the covariance 
and power counting indicate that multiplicative
renormalizability of the interacting theory may require introducing
two non-minimal terms linear in Ricci tensor in the vector sector.
Nevertheless, a more detailed analysis shows that these
non-minimal terms violate the decoupling of the longitudinal mode
and are prohibited. As a verification of general arguments, we derive
the one-loop divergences in the minimal massive scalar QED,
using St\"{u}ckelberg procedure and the heat-kernel technique. The
theory without non-minimal terms proves one-loop renormalizable
and admits the renormalization group equations for all the running
parameters in the scalar and vector sectors. One-loop beta functions
do not depend on the gauge fixing and can be used to derive the
effective potential.
\vskip 3mm

\vskip 2mm

\end{abstract}

\setcounter{footnote}{0} 
\renewcommand*{\thefootnote}{\arabic{footnote}} 

\section{Introduction}
\label{sec0}

The theory of a massive vector field  (Proca model) describes a
spin-1 massive vector particle. The mass term in the Lagrangian
distinguishes it from gauge vector field corresponding to massless
particles with helicities $\pm 1$.
Besides being the well-established subject of particle physics
and formal QFT considerations, there is currently a growing interest
in the study of Proca model in curved space-time, partially owing
to the new cosmological applications (see e.g.,
\cite{Ford-89,Kanno_2008,Golovnev_2008,Razieh_2017} and many
references therein). One can also mention the cosmological models
with the modified or generalized massive vector field actions (see,
e.g., \cite{Heisenberg_14,Heisenberg_16,Peter_16}). Also, for earlier
applications of this model in the gravitational field, see
\cite{Novello_79,Davies-Toms_85,Dimopoulos_2008}.

Exploration of a theory at the quantum level is relevant, not only
because of the quantum corrections to the classical action. At least
equally important is that the consistency of quantum field theory
enables one to restrict the form of the classical action. In this
respect, massive vector field $A_\mu$ and its extensions to the
curved spacetime attracted some attention. In particular,
publications \cite{Toms-2015,BST,Stein} were devoted to the
formulation of the theory in curved spacetime and evaluation of
vacuum quantum effects. The problem to solve in these papers turned
out to be very difficult and the corresponding calculations produced
conflicting results. The reason was that, in these works, the free
curved-space Proca model was formulated in a general form, with two
nonminimal terms $R_{\mu\nu} A^\mu  A^\nu$ and $RA^\mu  A_\mu$
included. On the other hand, one should question why it is necessary
to include these terms. Taking into account the well-known situation
with the nonminimal curvature-scalar field interaction (see, e.g.,
\cite{OUP}), the answer requires careful analysis of an interacting
theory.

An important feature of field theories in
curved space-time is the possibility of a non-minimal coupling of the
fields to gravity. A well-known example of such a coupling is the
famous $\xi R \varphi^2$ term in the scalar field Lagrangian. The
need for this term becomes clear in the framework of interacting
quantum theory, where the nonminimal term is a necessary element
of renormalizable theory \cite{OUP}).
One can ask whether the same situation takes place in the theory
of a curved-spacetime massive vector field coupled to matter. In
principle, the interaction with matter may generate the nonminimal
divergences of the form $R_{\mu\nu}A^{\mu}A^{\nu}$ and
$RA^{\mu}A_{\mu}$ in the vector sector.
In this case, the renormalizability would require us to introduce
these non-minimal terms to the classical action as an ultraviolet
completion.

The equations of motion in the free theory of massive vector
field $A_{\mu}$ yield the condition $\partial_{\mu}A^{\mu}=0$,
which means only transverse filed $A^{\bot}_{\mu}$ is propagating,
while the longitudinal mode (equivalent to a scalar filed) is
non-physical. Therefore, including new interactions of the
massive vector field to dynamical or external fields should not
lead to a propagating longitudinal mode at both classical and quantum
levels. This requirement imposes strong restrictions on the form of
possible interaction. In particular, this condition restricts possible
counterterms in the theory and is operational in ruling out the
non-minimal terms that are formally acceptable by power counting.

The renormalizability of massive vector field coupled to fermions
in flat space has been studied by many authors (see, e.g.,
\cite{BogShir} for earlier references). A full
analysis of renormalizability in flat space have been given in the
paper \cite{Boulware_70} where it was proved by the direct
transformations of the generating functional of Green functions
that the longitudinal mode of the vector field decouples
and the remaining theory of a transverse vector field is
renormalizable in power counting. Let us also mention recent
review \cite{Ruegg_2004} of the related subject, where one can
find further references.

In what follows, we prove the renormalizability of the Proca model
coupled to complex scalar field in curved space using two methods,
one is curved space generalization of the analysis considered in
\cite{Boulware_70} and second one is analysis of functional integral
for corresponding gauge theory with St\"{u}ckelberg field. The power
counting, in both approaches, indicates that the aforementioned
non-minimal terms are formally necessary. However, these non-minimal
terms violate the decoupling of a longitudinal mode of the vector
field and therefore are prohibited. The same output follows from the
analysis based on the gauge symmetry restored by means of the
St\"{u}ckelberg trick and the known 
gauge-fixing dependence in QFT. As a result, the multiplicative
renormalizability in curved space-time background does not require
us to take into account the curvature-dependent non-minimal terms.

The one-loop calculations in both free massive vector theory and
in the interacting models, can be performed by using the covariant
Schwinger-DeWitt technique (see e.g., \cite{DeWitt,bavi85}
and also the textbook \cite{OUP} for detailed introduction). The
application of this technique to the free vector field theory in
curved space can be found in \cite{bavi85} and \cite{BuGui} using
two different approaches, but with the equivalent results.
The  method of \cite{bavi85} is based on the introduction of an
auxiliary operator eliminating the degeneracy of the bilinear form
of the action \cite{bavi85}. The second approach relies on the
St\"{u}ckelberg procedure \cite{BuGui} (see also \cite{BST} and
\cite{Stein}), based on the introduction of an auxiliary scalar
field, restoring the gauge symmetry violated by the vector
mass. In this paper we follow the second approach, which we believe
is the most elegant and simple enough for direct loop calculations.

The rest of the paper is organized as follows. In Sec.~\ref{sec2},
we briefly review the scalar-vector model with a massive vector
field and introduce the  St\"{u}ckelberg procedure. Sec.~\ref{sec3}
presents the proof of the power-counting renormalizability of the
theory and completes the arguments in favor of the multiplicative
renormalizability of the minimal theory. Sec.~\ref{sec4} describes
the calculation of the one-loop divergences through the use of the
background field method,
St\"{u}ckelberg procedure and Schwinger-DeWitt technique. The main
text contains sufficient technical details, but part of the
intermediate formulas is separated in Appendix A. Furthermore, an
alternative calculational approach based on the auxiliary operator
and the subsequent difficulties are illustrated in Appendix B. The
elements of the renormalization group, i.e., the beta- and
gamma-functions are analyzed in Sec.~\ref{sec5}, where we also write
down the expression for the effective potential of the scalar.
Finally, in Sec.~\ref{sec6}, we draw our conclusions and present the
discussion of the validity of the one-loop results in the theory
under discussion.

\section{Curved-space scalar electrodynamics with massive vector field}
\label{sec2}

Let us start from the general formulation of massive vector field theory
coupled to a complex scalar and external gravitational field.
Unlike the conventional Abelian gauge-invariant scalar
electrodynamics, the theory under consideration is not a gauge
invariant theory owing to the mass of the vector.

\subsection{Classical action and notations}
\label{sec21}

The action of massive scalar electrodynamics, minimally coupled to
gravity in the vector sector, is written in the form
\beq
S[A,\Phi^{*},\Phi]
\,=\,\int\! d^{4}x\sqrt{-g}
\left\{
\, -\dfrac{1}{4}F_{\mu\nu}^{2}
-\dfrac{1}{2}M_v^2A_{\mu}^{2}
+ \left( \mathcal{D}^{\mu}\Phi\right)^{*}
\left(\mathcal{D}_{\mu}\Phi\right)
- V\left(\Phi^{*}\Phi\right)\right\}
\,\,\,
\label{eq:2.1}
\eeq
where
\beq
V\left(\Phi^{*}\Phi\right)
= m^2 \Phi^*\Phi + \la\left(\Phi^*\Phi\right)^2 - \xi R\Phi^*\Phi\,,
\label{V}
\eeq
$F_{\mu\nu}^{2}=F_{\mu\nu}F^{\mu\nu}$, $
F_{\mu\nu}=\partial _{\mu}A_{\nu}-\partial_{\nu}A_{\mu}$, and
$A_\mu^2=A_\mu A^\mu$. The mass of vector field is $M_v$, the mass
of the scalar field is $m$ and $\xi$ is the parameter of nonminimal
scalar-curvature interaction, which is well-known to be relevant in curved
space-time \cite{OUP}. Furthermore, the covariant derivatives are
\beq
\mathcal{D}_{\mu}\Phi = \na_\mu \Phi - ieA_{\mu}\Phi ,
\qquad
\big(\mathcal{D}_{\mu}\Phi\big)^* = \na_\mu \Phi^* + ieA_{\mu}\Phi^* .
\label{covder}
\eeq
Coupling constants include scalar self-interaction $\la$ and $e$.

\subsection{Reformulation using the St\"uckelberg procedure}
\label{sec22}

The peculiarity of the theory under consideration is that although
the vector sector of the theory (\ref{eq:2.1}) as a whole is not
gauge invariant, its kinetic term corresponds to the gauge-invariant
theory, which creates some difficulties in quantum calculations. In
particular, the covariant Schwinger-De Witt technique for the
effective action in external gravitational field can not be applied
directly. For a free massive vector theory in an external gravitational
field, these difficulties were overcome by the two different methods
in \cite{bavi85,BuGui}.
Trying to generalize these two methods to the interacting model
of massive vector and scalar (\ref{eq:2.1}), we met the following
situation. Different from the
pure vector case, the two methods do not give the same result in
the presence of a scalar field. One of the calculational procedures
is described in the next section and the difficulties that emerge in
another method are illustrated in the Appendix B.

The St\"uckelberg procedure introduces an additional real scalar
field $\th$ according to
\beq
A_\mu
\,\,\,
\longrightarrow
\,\,\,
A_{\mu} - \dfrac{1}{M_v} \na_\mu \th .
\eeq
In this way, the action of the theory becomes
\beq
S'[A,\theta,\Phi^{*},\Phi]
&=&
\int \! d^{4}x\sqrt{-g}\,\bigg\{
\!-\dfrac{1}{4}F_{\mu\nu}^{2}
- \dfrac{1}{2}{M_{v}^{2}}\Big(A_{\mu}-\dfrac{1}{M_{v}}
\na_{\mu}\th\Big)^2
\nn
\\
&&
+ \,\, g^{\mu\nu}\left(D_{\mu}\Phi\right)^{*}\left(D_{\nu}\Phi\right)
    -V\left(\Phi^{*}\Phi\right)\bigg\}
    \!\! ,
    \,\,\,\,\,\,
\label{eq:4.2}
\eeq
The last action is invariant under the gauge transformations
\beq
&&
\Phi\longrightarrow\Phi'\,=\,e^{ie\zeta(x)}\Phi,
\qquad\quad
\Phi^{*}\longrightarrow\Phi'^{*}\,=\,e^{-ie\zeta(x)}\Phi^{*},
\nn
\\
&&
A_{\mu}  \longrightarrow A'_{\mu}\,=\,A_{\mu}+\na_{\mu}\zeta(x),
\quad
\textrm{and}
\,\,\quad
\th\longrightarrow\th'\,=\,\th+M_{v}\zeta(x)\,.
\label{gaugetrans}
\eeq
Using the gauge invariance of the above action, we can impose the
gauge condition $\th=const$ which recovers the original action
(\ref{eq:2.1}). This shows to which extent the theories (\ref{eq:2.1})
and (\ref{eq:4.2}) are classically equivalent. The advantage of
the theory (\ref{eq:4.2}) is that it possesses the gauge invariance
and enables one to apply of the corresponding quantization scheme.

The equations of motion for all fields of the action (\ref{eq:4.2})
have the form
\beq
&&
\mathcal{E}^{*}
\,=\,\mathcal{D}_{\mu}\big(\mathcal{D}^{\mu}\Phi\big)^{*}
-\big (m^{2} -\xi R+2\la \Phi^{*}\Phi\big)\Phi^{*}\,=\,0,
\nn
\\
&&
\mathcal{E} 
\,=\,\mathcal{D}^{\mu}\big(\mathcal{D}_{\mu}\Phi\big)
- \big( m^2  - \xi R  + 2\la \Phi^*\Phi \big)\Phi \,=\,0,
\nn
\\
&&
\mathcal{E_{\theta}}
\,=\,-\na^{\mu}\na_{\mu}\th+M_{v}\na_{\mu}A^{\mu} \,=\,0,
\nn
\\
&&
\mathcal{E}^\nu
\,=\, \pa_\mu F^{\mu\nu} - M_{v}^{2}\Big(A^{\nu}
-\dfrac{1}{M_{v}}\na^{\nu}\th\Big) + J^\nu \,=\,0,
\label{eqPhiA}
\eeq
where $J^\nu = ie \big[ \Phi^* \mathcal{D}^\nu\Phi
- \Phi (\mathcal{D}^\nu\Phi)^* \big]$.

It proves useful to consider a linear real combination of
Eqs.~(\ref{eqPhiA}) with arbitrary constants $\al_1$,
$\,\al_2$, and $\al_3$,
\beq
&&
\dfrac{\al_1}{2} \big(\Phi^* \mathcal{E} +  \Phi \mathcal{E}^{*}\big)
+ \al_2\mathcal{E}_\th + \al_3 A_\nu \mathcal{E}^\nu
\,\,=\,\, \dfrac{\al_1}{2} \big[
  \Phi^*\mathcal{D}^{\mu}\big(\mathcal{D}_{\mu}\Phi\big)
+ \Phi\mathcal{D}^{\mu}\big(\mathcal{D}_{\mu}\Phi\big)^{*}
\big]
\nn
\\
 &&
- \,\, \al_1 \big( m^2  - \xi R\big)\Phi^*\Phi
\, - \, 2\al_{1}  \la  \big(\Phi^*\Phi \big)^2
- \al_2 \th \na^\mu \na_\mu \th
+ \al_2 M_v \th \na_\mu A^\mu
\nn
\\
 &&
+ \,\,\al_3 A_\nu \pa_\mu F^{\mu\nu}
\,- \,
 \al_3 M_v^2 A^2_\nu
+ \al_3 M_{v}A_{\nu}\na^{\nu}\th
+  ie \al_3 A_\nu \big[ \Phi^* \mathcal{D}^\nu\Phi
- \Phi (\mathcal{D}^\nu\Phi)^* \big].
 \qquad
\label{TracePhiA}
\eeq
An equivalent expression will be used in Sec.~\ref{sec4} for
identifying essential effective charges in the renormalization
group running.

\section{Power counting and renormalization in curved space-time}
\label{sec3}

In this section, we generalize the analysis of power counting made
in the paper \cite{Boulware_70} for massive vector minimally coupled
to fermions in flat space (see also the review \cite{Ruegg_2004}), for
the theory in curved space-time. Our main goal is to formulate the
curved-space action which guarantees the renormalizability of the
theory of massive vector field coupled to charged scalars. Similar
to \cite{Boulware_70}, we show that the theory under consideration
is renormalizable by power counting, but also highlight new aspects
of its renormalization in curved spacetime. The analysis is carried
out in two ways: within the framework of a model with an auxiliary
St\"uckelberg field (\ref{eq:4.2}) and directly using the original form
of the model (\ref{eq:2.1}). The results regarding renormalizability
obviously coincide, which is not surprising, since the models are
classically equivalent and there are no sources of quantum anomalies.
The consideration can be easily extended including the interaction
to Dirac fermions or to both scalars and fermions in curved
spacetime.

\subsection{Model with auxiliary field}
\label{sec31}

Our starting point will be the gauge invariant action (\ref{eq:4.2}).
The gauge transformations have the form (\ref{gaugetrans}). We begin
with the generating functional of the Green functions
\beq
&&
Z[J_{\mu},J,J^{*}]
\,=\, \int {\cal D}A \,{\cal D}\theta\,{\cal D}\,\Phi^{*}\,{\cal D}\,
\Phi \,\,\delta[\chi]                                                                                                          \,\,
\nonumber
\\
&& \times \,\,
\exp \Big\{
i S'[A,\theta,\Phi^{*},\Phi]
+ i \int d^4x \sqrt{-g} \big(A_{\mu}J^{\mu}
+ \Phi^{*}J + \Phi J^{*}\big) \Big\},
\eeq
where $\chi$ is a gauge fixing function.
Now we present the vector field as a sum of transverse and
longitudinal parts, $A_{\mu}=A^{\bot}_{\mu} + \partial_{\mu}\ph$,
where $\nabla^{\mu}A^{\bot}_{\mu}=0$.
Then the action takes the form
\beq
\nonumber
&&
S'[A^{\bot},\ph,\theta,\Phi^{*},\Phi]
\,=\, \int d^4x \sqrt{-g} \,\,
\Big\{
- \frac{1}{4}F^{\bot}_{\mu\nu}F^{\bot\mu\nu}
+ \frac{1}{2} M_v^2 g^{\mu\nu}
\Big(A^{\bot}_{\mu}+\partial_{\mu}\ph
- \frac{1}{M_v}\partial_{\mu}{\theta}\Big)
\\
\nonumber
&&
\qquad
\times\,
\Big(A^{\bot}_{\nu}+\partial_{\mu}\ph
- \frac{1}{M_v}\partial_{\nu}{\theta}\Big)
+ g^{\mu\nu}\big(\partial_{\mu} + ieA^{\bot}_{\mu}
+ ie \partial_{\mu}\ph\big)\Phi^{*}
\big(\partial_{\nu} - ieA^{\bot}_{\nu}
- ie\partial_{\mu}\ph \big)\Phi
\\
&&
\qquad
\quad
- \,\,V\left(\Phi^{*}\Phi\right) \Big\},
\label{newact2}
\eeq
where the classical scalar potential $V\left(\Phi^{*}\Phi\right)$ is
given by (\ref{V}). For further consideration, it is convenient to
fulfill a change of variables in the functional integral,
\ $A_{\mu} \rightarrow (A^{\bot}_{\mu}, \varphi)$, with
$F^{\bot}_{\mu\nu} = F_{\mu\nu}$.
This is a linear change of variables, hence the corresponding Jacobian
depends only on external metric and does not depend of the fields
$A^{\bot}_{\mu}, \varphi$. Since we are interested in the
renormalization of the matter field sector of the effective action (see, e.g., \cite{BuGui}), this Jacobian will
be omitted in the rest of this paper.\footnote{At one loop, the
vacuum functional is a sum of the contributions of free massive
vector \cite{bavi85,BuGui} given Eq.~(\ref{vacu}) and of the scalar
field. Thus, its special evaluation in the interacting theory is not
relevant.} Then the generating functional takes the form
\beq
\nonumber
&&
Z[J_{\mu},J,J^{*}]
\,\,=\,\,
\int {\cal D}A^{\bot}\,{\cal D} {\ph}\,{\cal D}\theta\,{\cal D}\Phi^{*}\,
{\cal D}\Phi \,\,\delta[\chi]
\\
&&
\qquad
\times \,\,
\exp\Big\{
iS'[A^{\bot},\ph,\theta,\Phi^{*},\Phi]
+ i \int d^4x \sqrt{-g}\,\big(
A_\mu J^\mu  + \Phi^{*}J + \Phi J^{*}\big)\Big\},
\label{Z2}
\eeq
where $\chi$ is the gauge fixing condition\footnote{The corresponding
Faddeev-Popov determinant depends only on the external metric and
hence it does not affect the renormalizability in the matter fields sector.
For this reason, it will be omitted.}
It is convenient to take such a condition in the form
\beq
\chi \,=\, \frac{\theta}{M_v}\, -\, \ph.
\eeq
Owing to the factor of \ $\delta \big(\chi\big)$ in (\ref{Z2}),
we get
\beq
g^{\mu\nu}\big(
A^{\bot}_{\mu}
+ \partial_{\mu}\ph
- \frac{1}{M_v}\partial_{\mu}{\theta}\big)
\big(A^{\bot}_{\nu}+\partial_{\nu}\ph
- \frac{1}{M_v}\partial_{\nu}{\theta}\big)
\,\,=\,\, g^{\mu\nu}A^{\bot}_{\mu}A^{\bot}_{\nu}.
\label{Shtu}
\eeq
As a result, we obtain the following action in the integrand:
\beq
&&
S'[A^{\bot},\Phi,\Phi^{*},\Phi]
\,=\, \int d^4x \sqrt{-g}
\,\,\Big\{
- \frac{1}{4}F_{\mu\nu}F^{\mu\nu}
+ \frac{1}{2} {M_v}^2 g^{\mu\nu}A^{\bot}_{\mu}A^{\bot}_{\nu}
\nonumber
\\
&&
\quad
+\,
g^{\mu\nu}\big(\partial_{\mu} + ieA^{\bot}_{\mu}
+ ie\partial_{\mu}\ph \big)\Phi^{*}
\big( \partial_{\nu} - ieA^{\bot}_{\nu} - ie \partial_{\mu}\ph)\Phi
\nonumber
\\
&&
\qquad
-\,\,V\left(\Phi^{*}\Phi\right)\Big\}.
\eeq

Next, we perform the following change of variables in the
functional integral:
\beq
\Phi \rightarrow e^{ie\ph}\Phi,
\qquad
\Phi^{*} \rightarrow e^{-ie\ph}\Phi^{*},
\label{changePhi}
\eeq
without changing the notations for the variables.
After this, the field $\ph$ disappears completely from the
action and we obtain the generating functional in the form,
\beq
&&
Z[J_{\mu},J,J^{*}]
\,\,=\,\,
\int {\cal D}A^{\bot}\,{\cal D} {\ph}\,{\cal D}\theta
\,{\cal D}\Phi^{*}\,{\cal D}\Phi\,\,
\delta \Big(\frac{\th}{M_v} - \ph\Big)
\nn
\\
&&
\qquad
\times \,\,
\exp\Big\{
iS'[A^{\bot},\theta,\Phi^{*},\Phi]
\,+\,
i \int d^4x \sqrt{-g}\,\big(
A_{\mu}J^{\mu} + \Phi^{*}J  +  \Phi J^{*}\big)\Big\}.
\label{Z3}
\eeq
Integrating over $\ph$ yields
\beq
&&
Z[J_{\mu},J,J^{*}]
\,\,=\,\,
\int {\cal D}A^{\bot}\,{\cal D}\theta\,{\cal D}\Phi^{*}\,{\cal D}\Phi \,\,
\exp\Big\{ iS' [A^{\bot},\theta,\Phi^{*},\Phi ]
\nn
\\
&&
\qquad \qquad \qquad
+ \,\,i\int d^4x \sqrt{-g}\,\Big(
A^{\bot}_{\mu}J^{\bot{\mu}}
+ \frac{1}{M_v} J^{\mu}\pa_\mu \th \,+\,\Phi^{*}J + \Phi J^{*}\Big)
\Big\}.
\qquad
\eeq
In the last expression, we introduced the transverse and longitudinal
sources according to $J^{\mu} = J^{\bot}_{\mu}+\partial_{\mu}j$,
where $\nabla^{\mu}J^{\bot}_{\mu}=0$. Disregarding total derivative,
$J^{\mu} \partial_{\mu}\theta = - \,\theta\Box j$.
Now we can integrate over $\theta$. According to the delta function
property, this operation gives
\beq
\delta[\Box j] \,=\, \frac{1}{Det\,{\cx}}\,\,\delta[j]\,.
\eeq
The expression $Det \,{\cx}$ depends only on the external metric
and can be ignored when analyzing the renormalizability in the
matter sector.

Thus, we arrive at the expression
\beq
&&
Z[J_{\mu},J,J^{*}]
\,=\, \delta[j]
\int {\cal D}A^{\bot}\,{\cal D}\Phi^{*}\,{\cal D}\Phi\,
\exp\Big\{
i[S[A^{\bot},\theta,\Phi^{*},\Phi]
\nn
\\
&&
\qquad \qquad \qquad
+ \,\,
i \int d^4x \sqrt{-g}(A^{\bot}_{\mu}J^{\bot{\mu}}+\Phi^{*}J
+ \Phi J^{*})\Big\}.
\label{Z4}
\eeq
We obtained the generating functional for the theory of the
fields $A^{\bot},\Phi^{*},\Phi$ with the action
\beq
&&
S'[A^{\bot},\theta,\Phi^{*},\Phi]
\,=\,  \int d^4x \sqrt{-g}\Big\{
- \frac{1}{4}F_{\mu\nu}F^{\mu\nu}
+ \frac{1}{2} M_v^2 \,g^{\mu\nu}A^{\bot}_{\mu}A^{\bot}_{\nu}
\nonumber
\\
&&
\qquad
+\,\,
g^{\mu\nu}(\partial_{\mu} \Phi^{*} + ieA^{\bot}_{\mu}\Phi^{*})\,
(\partial_{\nu}\Phi - ieA^{\bot}_{\nu}\Phi)
- V\left(\Phi^{*}\Phi\right) \Big\}.
\qquad
\label{newact}
\eeq

To analyze the superficial degree of divergences it is sufficient to
represent external metric in the form $g_{\mu\nu}=\eta_{\mu\nu}
+h_{\mu\nu}$ and expand all the terms containing $g_{\mu\nu}$
in power series in $h_{\mu\nu}$. All these terms can be treated
as the perturbations. In this case the propagators of the field
$A^{\bot}$ and of the fields $\Phi^{*},\Phi$ are, respectively,
\beq
\frac{1}{p^2 - M_v^2}
\,\Big(
\delta_{\mu}{}^{\nu}-\frac{p_{\mu}p^{\nu}}{p^2}\Big)
\qquad
\mbox{and}
\qquad
\frac{1}{p^2-m^2}.
\label{propaAtransPhi}
\eeq
After the Wick rotation, the behavior of both propagators at
$p \rightarrow \infty$ is of the $p^{-2}$ type. Since the field
$h_{\mu\nu}$ is dimensionless, its powers do not contribute to the
superficial degree of divergence $\omega$. Then the standard
estimate is
\beq
\omega \,=\, 4 - N_{A}-N_{\Phi},
\label{sdg-1}
\eeq
where $N_{A}$ is the number of  external lines of the field
$A^{\bot}$ and $N_{\Phi}$ is a number of external lines of the
fields $\Phi^{*},\Phi$. As a result, we conclude that the theory is
renormalizable by power counting. However, as usual in curved space,
this does not guarantee the multiplicative renormalizability of the
theory if only minimal coupling to gravity is considered (see e.g.,
\cite{OUP}). Let us analyze what are the possible
covariant and local counterterms fitting the estimate (\ref{sdg-1}).

First we consider the option with $N_{\Phi}=2$ and $N_{A}=0$. In
this case, the unique nonminimal counterterm is the one of the form
$R\Phi^{*}\Phi$, which is extensively discussed in the literature
(see e.g., \cite{OUP}). Let us note that this term is already included in the classical scalar potential $V\left(\Phi^{*}\Phi\right)$.
It is worth noting that including this term into the classical action
does not modify neither the arguments presented below nor the power
counting.

Let us concentrate on another possible option, with
$N_{\Phi}=0$ and $N_{A}=2$. In this case, $\omega=2$.
Hence, the counterterms in the sector of the fields $A^{\bot}$
may have the form
\beq
-\,\frac{1}{4}\delta z_1 F_{\mu\nu}F^{\mu\nu}
+ \delta z_2 \,\frac{1}{2} M_v^2 A^{\bot\mu}A^{\bot}_{\mu}
+ \frac{1}{2}\,\delta z_3 \,R A^{\bot\mu}A^{\bot}_{\mu}
+ \frac{1}{2} \delta z'_{3} \,R^{\mu\nu} A^{\bot}_{\mu}A^{\bot}_{\nu}.
\label{contaA}
\eeq
Owing to the locality and covariance of divergences, one can
safely replace $A^{\bot}_{\mu}$ by $A_{\mu}$ in the last
expression, hence we can expect, in the vector sector, the
counterterms of the form
\beq
F_{\mu\nu}F^{\mu\nu},
\qquad
M_v^2 A_\mu A^{\mu},
\qquad
R A_\mu A^\mu,
\qquad
R^{\mu\nu} A_{\mu}A_{\nu}.
\label{contrasA}
\eeq
The first two terms are present in the classical action with minimal
coupling to gravity in vector sector, which corresponds to the
renormalizability by power counting. On the
other hand, the last two terms are non-minimal and were not
included into the initial Lagrangian. According to the conventional
approach used in case of scalars (and some other cases, e.g., the
theory with torsion \cite{bush85} or other external fields
\cite{CPTL-Gui}, to achieve the renormalizable theory we
have to include into classical action these non-minimal terms with
the coupling of massive Abelian vector field to the external gravity.
However, the present case is different. We leave the
discussion of these terms to the last part of this
section, after an additional analysis of the power counting by
a different method.

Finally, there is one more type of terms, with $N_{\Phi}=0$ and
$N_{A}=4$, such that $\omega=0$. In this case, the counterterms
in the sector of the fields $A^{\bot}$ should have the form
$\delta z_4 \, (A^{\bot\, \mu}A^{\bot}_{\mu})^2$. In the covariant
local form, the corresponding counterterm is $(A^{\mu}A_{\mu})^2$.
In principle, the multiplicative renormalizability requires us to
include this kind of a term into initial classical action, already in
flat spacetime. As we know from the analysis (including explicit
two-loop calculations) of axial vector model
\cite{guhesh}, this term may result in the longitudinal divergences
of the form $(\pa_\mu A^{\|\, \mu})^2$ in higher loop orders. In its
turn, this means the violation of unitarity at the quantum level
\cite{SlavFadd}. These arguments were operational and effectively
used in \cite{Beltors} for constructing the action of a propagating
axial vector dual to antisymmetric torsion. Power counting alone
cannot provide the protection against this scenario, so it requires
an additional detailed consideration. We shall see at the last part
of this section,  that in the case of the curved-space Proca model,
there is a protection against this scenario.

\subsection{The model without auxiliary field}
\label{sec32}

For the sake of completeness, let us consider the power counting
without making the St\"{u}ckelberg trick. This approach is quite
close to the one of \cite{Boulware_70}. The starting action is (\ref{eq:2.1})
and the generating functional of Green functions has the form
\beq
&&
Z[J_{\mu},J,J^{*}]
\,=\,
\int {\cal D}A\,{\cal D}\th\,{\cal D}\Phi^{*}\,{\cal D}\Phi \,\,
\exp\Big\{
iS[A,\Phi^{*},\Phi]
\nn
\\
&&
\qquad\qquad\qquad
+ \,\, i\int d^4x \sqrt{-g}\,\big(
A_{\mu}J^{\mu} + \Phi^{*}J + \Phi J^{*}\big)\Big\}.
\label{Z5}
\eeq
As in the previous version of the proof, one can make the change
of variables $A_\mu = A^\bot_\mu + \pa_\mu \ph$. The mass term
in the vector sector transforms as follows:
\beq
\nonumber
&&
g^{\mu\nu}A_{\mu}A_{\nu}
\,=\,  g^{\mu\nu}(A^{\bot}_{\mu}
+ \partial_{\mu}\ph)(A^{\bot}_{\nu}+\partial_{\nu}\ph)
\nn
\\
&&
\qquad
= \,\, g^{\mu\nu}A^{\bot}_{\mu}A^{\bot}_{\nu}
\,+\, g^{\mu\nu}\,\pa_\mu \ph \,\pa_\nu \ph
\,+\, \mbox{total derivatives}.
\eeq

After the change of variables (\ref{changePhi}) in the scalar
sector, we arrive at the action
\beq
\nonumber
&&
S[A^{\bot},\ph,\Phi^{*},\Phi]
\,=\,
\int d^4x \sqrt{-g}\,\Big\{
- \,\frac{1}{4}F^{\bot}_{\mu\nu}F^{\bot\mu\nu}
+ \frac{M_v^2}{2}\,g^{\mu\nu}A^{\bot}_{\mu}A^{\bot}_{\nu}
+ \frac{M_v^2}{2}g^{\mu\nu}\,\pa_\mu \ph \pa_\nu \ph
\\
\nonumber
&&
\qquad
+ \,g^{\mu\nu}\big(\partial_{\mu}+ieA^{\bot}_{\mu}\big)\Phi^{*}
\big(\partial_{\nu} - ieA^{\bot}_{\nu}\big)\Phi
\,-\, m^2\Phi^{*}\Phi -\lambda(\Phi^{*}\Phi)^2 \Big\}.
\label{actnew}
\eeq

The generating functional becomes
\beq
&&
Z[J^{\bot}_{\mu},j,J,J^{*}]
\,=\, \int {\cal D}A\,{\cal D}\theta \, {\cal D}\ph\, {\cal D}\Phi^{*}
\, {\cal D}\Phi \,
\exp \big\{
i S[A^{\bot},\ph,\Phi^{*},\Phi]
\nn
\\
&&
\qquad
\qquad
+ \,\,i \int d^4x \sqrt{-g}
\,(A^{\bot}_{\mu}J^{\bot\mu}-\ph\Box j +\Phi^{*}J + \Phi J^{*})]
\big\}.
\eeq

The integral over $\ph$ gets factorized in form of $Z_2[j]$ and does not depend
of $A^{\bot}$ with
\beq
\nonumber
&&
Z_2[j]
\,\,=\,\,
\int {\cal D}\ph \exp
\Big\{
i\int d^4 x \sqrt{-g}
\,\Big(-\frac{M_v^2}{2}\,\ph\Box \ph - \varphi\Box j \Big)\Big\}
\nn
\\
&&
\qquad
= \,\,
\big[ Det \,\cx \big]^{- 1/2}
\exp\Big\{
\,\frac{i}{2M_v^2}\int d^4x \sqrt{-g}\,\,j{\cx} j\Big\}.
\label{intph}
\eeq
As a result, we get
\beq
Z[J^{\bot}_{\mu},j,J,J^{*}]
\,\,=\,\,Z_1[J^{\bot}_{\mu},J,J^{*}]Z_{2}[j],
\label{Z6}
\eeq
where $Z_1[J^{\bot}_{\mu},J,J^{*}]$ is the generating functional for
the theory with the action
\beq
\nonumber
&&
S[A^{\bot},\Phi^{*},\Phi]
\,=\, \int d^4x \sqrt{-g}\,\Big\{
- \frac{1}{4}F_{\mu\nu}F^{\mu\nu}
+ \frac{M_v^2}{2} g^{\mu\nu} A^{\bot}_{\mu}A^{\bot}_{\nu}
\\
&&
\qquad
+\,\, g^{\mu\nu}(\partial_{\mu} + ie A^{\bot}_{\mu})\Phi^{*}
(\partial_{\nu} - ieA^{\bot}_{\nu})\Phi
- m^2\Phi^{*}\Phi -\lambda(\Phi^{*}\Phi)^2 \Big\}.
\eeq

It is easy to see that all relevant divergences are concentrated only
in $Z_1[J^{\bot}_{\mu},J,J^{*}]$. Further considerations are
the same as in the previous section.

\subsection{Renormalizability of the minimal vector theory}
\label{sec33}

Now we are in a position to discuss
whether the divergences of the nonminimal form $RA^{\mu}A_{\mu}$
and $R^{\mu\nu} A_\mu A_\nu$, and of the self-interacting form
$\big(A^{\mu}A_{\mu}\big)^2$, can be expected at one- or
higher-loop orders.
At the one-loop level, in the Abelian theory, the diagrams with two
external vector lines are those shown in Fig.~1. These diagrams are
exactly the same as in the scalar QCD and, therefore, they will
preserve the gauge invariance in the counterterms. This feature rules
out the non-minimal terms since these term violate the symmetry.
\begin{figure}[htb]
\begin{center}
$\,$\includegraphics[angle=0,width=8.6cm]{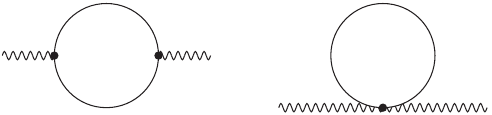}$\,$
\end{center}
\begin{quotation}
\caption{Diagrams contributing to bilinear vector terms
in the one loop order. }
\end{quotation}
\label{Fig1}
\end{figure}

The main reason of why the one-loop diagrams preserve gauge
invariance is that there are no internal massive vector lines,
violating the gauge invariance. Starting from the second loop,
the situation is different.  In the two-loop diagram shown in
Fig.~2, there is such an internal  line and, therefore, according
to the power counting, the nonminimal terms look possible
in the second- or higher-loop orders.

\begin{figure}[H]
\begin{center}
$\,$\includegraphics[angle=0,width=4.0cm]{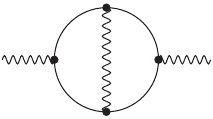}$\,$
\end{center}
\begin{quotation}
\caption{One of the higher-loop diagrams expected to
generate nonminimal term.}
\end{quotation}
\label{Fig2}
\end{figure}
\vspace{-1 cm}

The situation is quite similar in the case of
$\big(A^{\mu}A_{\mu}\big)^2$, so we skip the details of this case.
The final result is that this type of divergences may be expected at
higher loops if we follow the dimensional arguments, without
taking into account the symmetries.

Let us show that these arguments are not valid and, in fact, the
minimal theory (\ref{eq:2.1}), without introducing the new
curvature-dependent or $\big(A^{\mu}A_{\mu}\big)^2$ terms, is
all-loop renormalizable. The  starting point of our consideration
will be the symmetric formulation of the action (\ref{eq:4.2}).
Then the general QFT
theorems (see, e.g., \cite{Lavrov-Curved} and references therein)
tell us that, even in curved spacetime, the gauge symmetry under
(\ref{gaugetrans}) holds in the counterterms and, on top of this,
the counterterms are local.

The non-minimal and self-interacting terms in the
original version (\ref{eq:2.1}), have the form
\beq
\De S[A,\Phi^{*},\Phi]
\,=\,\int\! d^{4}x\sqrt{-g}
\left\{
\, \dfrac{1}{2} \ze_1 R_{\mu\nu} A^\mu A^\nu
\,+ \, \dfrac{1}{2} \ze_2 R A^\mu A_\mu
\,-\, \dfrac{f}{4!} (A^\mu A_\mu)^2
\right\}
\,\,\,
\label{eq:2.1extra}
\eeq
On the other hand, the symmetric version of these terms, in the
formulation (\ref{eq:4.2}), are more complicated,
\beq
&&
\De S[A,\Phi^{*},\Phi]\bigg|_{Stuck}
\,=\,\int\! d^{4}x\sqrt{-g}
\bigg\{
\, \dfrac{1}{2} \ze_1 R^{\mu\nu}
\Big( A_{\mu} - \dfrac{1}{M_v} \na_\mu \th\Big)
\Big( A_{\nu} - \dfrac{1}{M_v} \na_\nu \th\Big)
\nn
\\
&&
\qquad
\,\,+ \, \dfrac{1}{2} \ze_2 R
\Big( A_{\mu} - \dfrac{1}{M_v} \na_\mu \th\Big)^2
- \,\,\dfrac{f}{4!} \bigg[
\Big( A_{\mu} - \dfrac{1}{M_v} \na_\mu \th\Big)
\Big( A^{\mu} - \dfrac{1}{M_v} \na^\mu \th\Big)
\bigg]^2
\bigg\}.
\qquad
\label{eqextra}
\eeq
The main observation is that
this expression is incompatible with the power counting of the
theory, as it was established above. The point is that there are
terms with the inverse powers of mass $M_v$ and these factors
are compensated by the extra derivatives of the scalar field $\th$.
These terms are typical for the nonrenormalizable by power
counting theory, but this is in contradiction with the evaluation
presented above, in subsection \ref{sec31}. We conclude that
the counterterms of the form (\ref{eqextra}) cannot emerge in
the theory with the restored gauge symmetry (\ref{eq:4.2}).

Does the last conclusion apply to the original theory (\ref{eq:2.1})?
To answer this question we note that this theory corresponds to
(\ref{eq:4.2}) under the particular gauge fixing. On the other hand,
the general QFT theorems \cite{aref,Costa-Tonin,Kummer,BLT-YM2}
tell us that the difference between the divergences in two different
gauges is always proportional to the equations of motion. In our
case, the divergences in the theory (\ref{eq:2.1}) differ from the
ones in the theory (\ref{eq:4.2}) only by the terms proportional to
the equations of motion of the theory with quantum corrections.

One can use this information to prove renormalizability. To this end,
one can use iterations method. Taking into account the locality of
divergences and assuming multiplicative renormalizability of the
minimal theory (i.e., starting from the action without
(\ref{eq:2.1extra}) terms) at the $n$-th loop order, we can see that the
corresponding equations of motion form the same combination
(\ref{TracePhiA}) as we met at the classical level. The only change
concerns the coefficients $\al_{1,2,3}$ in this combination. In
view of the form of Eqs.~(\ref{eqPhiA}), we conclude that the terms
 (\ref{eq:2.1extra}) are ruled out as divergences in the original
 theory (\ref{eq:2.1}) in all loop orders.

There are also more simple arguments leading to the same conclusion.
Let us decompose the vector field in transverse and longitudinal
parts according to
$A_{\mu}=A^{\bot}_{\mu} +\, \pa_\mu\varphi$
and replace it into (\ref{eq:2.1extra}). In this way, we get
\beq
&&
\De S[A]
\,=\,\int\! d^{4}x\sqrt{-g} \,\,\Big\{
\, \dfrac{1}{2} \ze_1 R^{\mu\nu}
(A^{\bot}_{\mu}A^{\bot}_{\nu}+\partial_{\mu}\varphi\partial_{\nu}\varphi
+ 2A^{\bot}_{\mu}\partial_{\nu}\varphi)
\nn
\\
&&
\qquad
\,+ \, \,
\dfrac{1}{2} \ze_2 R( A^{\bot \,\mu} A^{\bot}_{\mu}
+ \na_{\mu}\varphi\nabla^{\mu}\varphi)
\,+\,\dfrac{f}{4!} \big(A^{\bot}_{\mu}A^{\bot \mu}
+ \na_\mu \varphi \na^\mu \varphi
+ 2A^{\bot}_{\mu} \na^\mu \varphi \big)^2 \Big\} .
\qquad
\label{eq:2.1superextra}
\eeq
It is clear that the longitudinal mode $\varphi$ in the terms containing
$R_{\mu\nu}$ and $A^4$ does not decouple from transverse
 component \ $A^{\bot}$. On the other hand, in the subsection
 \ref{sec32} it was proved 
 that the non-physical longitudinal mode
decouples in minimal theory.
This means, starting with the minimal theory, the aforementioned
nonminimal terms cannot be generated. The only term in
(\ref{eq:2.1superextra}) where the longitudinal mode decouples from
the transverse vector is the one with the coefficient $\ze_2$.

To understand the difference between the terms with $M_{v}A^2$
and $RA^2$, we note that for the massive term the decoupling of
transverse and longitudinal modes is described by the expression
\beq
\frac{1}{2}\, M^ 2_{v}A_\mu A^\mu
\,=\, \frac{1}{2}\, M^ 2_{v} \big[A^\bot_\mu A^{\bot \mu}
+ \na_\mu \varphi \nabla^\mu\varphi\big],
\label{Mepa}
\eeq
while for the curvature-dependent part we get
\beq
\frac{1}{2}\, \ze_2 R A_\mu A^\mu
\,=\, \frac{1}{2}\, R \big[A^\bot_\mu A^{\bot \mu}
+ \na_\mu \varphi \nabla^\mu\varphi\big].
\label{Repa}
\eeq
It is necessary to make the rescaling $\chi = M_v\ph$, such that the
new scalar $\chi$ gains the canonical dimension. After this, the sum
of the scalar sectors in (\ref{Mepa}) and (\ref{Repa}) becomes
\beq
\frac{1}{2}\, \Big( 1\,+\,\frac{ \ze_2 R}{M^ 2_v}\Big)
\na_\mu \chi \nabla^\mu\chi.
\label{RepaMepa}
\eeq
In case of $\ze_2 \neq 0$, this relation contains the inverse mass
coefficient that cannot be obtained as a divergence in the minimal
theory because of the power counting arguments. Therefore, the
parameter $\ze_2$ must be equal to zero. As a result, we see that
all the non-minimal terms in the vector sector are forbidden. We
can state that the minimal theory is multiplicative renormalizable,
including in the curved space-time.

\section{St\"uckelberg procedure and one-loop divergences}
\label{sec4}

In this section we perform calculation of the one-loop
divergences in the interacting theory. The practical calculations
are possible when using the St\"uckelberg trick, as it was
suggested in the free Proca model in \cite{BuGui}. Thus,
for the sake of covariant one-loop calculations, we will work
with the gauge invariant action $S'[A,\theta,\Phi^{*},\Phi]$
defined in (\ref{eq:4.2}). This action contains the St\"uckelberg
field $\theta$. Quantization is carried out using the Faddeev-Popov
Ansatz, involving the use of gauge-fixing term and the ghost action.

The useful form of the covariant gauge-fixing action is
\beq
S_{\textrm{gf}}
\,  = \, - \,\dfrac{1}{2}\int d^{4}x\sqrt{-g}\,\chi^{2},
\quad\textrm{where}\quad
\chi=\na_{\mu}A^{\mu}-M_{v}\th\,.
\label{eq:4.3}
\eeq
It is worth noting that this gauge condition is different from the
$\th = const$ that provides the equivalence of  the original theory
(\ref{eq:2.1}) and the theory with restored gauge symmetry
(\ref{eq:4.2}). Of course, such a gauge is inconvenient for loop calculations. Luckily, there are general arguments that enable
one to take into account the gauge-fixing dependence of the result.
We shall present these arguments in the end of this section. This
part is especially relevant because an alternative method of the
heat-kernel calculations, as described in Appendix B, is difficult
to apply in the present case.

The sum of the action  (\ref{eq:4.2}) and the gauge-fixing
 term (\ref{eq:4.3}) has the form
\beq
S'+S_{\textrm{gf}}
&=&
\int d^{4}x\sqrt{-g}\,\bigg\{ \dfrac{1}{2}A^{\nu}
\left(\de_{\nu}^{\mu}\cx-R^{\mu}{}_{\nu}
- \de_{\nu}^{\mu}M_v^2\right) A_{\mu}
+ \dfrac{1}{2}\, \th\left(\cx-M_{v}^{2}\right)\th
\nn
\\
&&
+\,\, g^{\mu\nu}
\left(D_{\mu}\Phi\right)^{*}\left(D_{\nu}\Phi\right)
- m^{2}\Phi^{*}\Phi
- \la\left(\Phi^{*}\Phi\right)^2
+ \xi R\Phi^{*}\Phi \bigg\} .
\label{eq:4.4}
\eeq

To apply the background field method, we
decompose the fields into classical and quantum
$(\ph, \,\ph^*, \,B^\mu,\,\te)$ counterparts,
\beq
    \Phi^{*}\rightarrow\Phi^{*}=\Phi^{*}+\ph^{*},
    \quad\,
    \Phi\rightarrow\Phi=\Phi+\ph,
    \quad\,
    A_{\mu}\rightarrow A_{\mu}=A_{\mu}+B_{\mu}
\quad\,
    \th\rightarrow\th=\th+\te
\eeq
and extract the bilinear in quantum fields form of the
action (\ref{eq:4.4})
\beq
S'^{(2)}+S_{\textrm{gf}}
&=&
\dfrac{1}{2}\int d^{4}x\sqrt{-g}
\begin{pmatrix}
    \ph^{*} & \ph & \te & B^{\mu}
    \end{pmatrix}
    \hat{H'}
    \begin{pmatrix}
        \ph\\
    \ph^{*}\\
    \te\\
    B_{\nu}
\end{pmatrix}\,.
\eeq
The differential operator $\hat{H'}$ has a matrix form
\beq
\label{stheis}
\hat{H}'
&=&
\begin{pmatrix}
\label{eq:4.7}
\hat{H}'_{\ph^{*}\ph} & \hat{H}'_{\ph^{*}\ph^{*}}
& \hat{H}'_{\ph^{*}\te}
& \hat{H}'_{\ph^{*}B_{\nu}}    \\
    \hat{H}'_{\ph\,\ph}
& \hat{H}'_{\ph\,\ph^{*}}
& \hat{H}'_{\ph\,\te}
& \hat{H}'_{\ph\, B_{\nu}}    \\
    \hat{H}'_{\te\,\ph}
& \hat{H}'_{\te\,\ph^{*}}
& \hat{H}'_{\te\,\te}
& \hat{H}'_{\te\, B_{\nu}}       \\
    \hat{H}'_{B^{\mu}\ph}
& \hat{H}'_{B^{\mu}\ph^{*}}
& \hat{H}'_{B^{\mu}\te}
& \hat{H}'_{B^{\mu}B_{\nu}}\,,
\end{pmatrix}
\eeq
with the following non-vanishing elements:
\beq
\label{A7}
&&
\hat{H}'_{\ph^{*}\ph\,}
\;=\;
-\cx-m^{2}
-4\la\Phi^{*}\Phi
+\xi R
+2ieA_{\mu}\na^{\mu}
+ie\left(
\na_{\mu}A^{\mu}
\right)
+e^{2}A^{2}_{\mu},
 \nn
\\
&&
\hat{H}'_{\ph\,\ph^{*}}
\;=\;
-\cx
-m^{2}
-4\la\Phi^{*}\Phi
+\xi R
-2ieA_{\mu}\na^{\mu}
-ie\left(
\na_{\mu}A^{\mu}
\right)
+e^{2}A^{2}_{\mu} ,
\nn
\\
&&
\hat{H}'_{B^{\mu}B_{\nu}}
\!=\;
\de_{\mu}^{\nu}\cx -R_{\mu}{}_{.}^{\nu}
+\de_{\mu}^{\nu}M_{v}^{2}
+2\de_{\mu}^{\nu}e^{2}\Phi^{*}\Phi,
\nn
\\
&&
\hat{H}'_{\ph^{*}B_{\nu}}
\,=\;
ie\Phi\na^{\nu}
+2ie\left(
\na^{\nu}\Phi
\right)
+2e^{2}A^{\nu}\Phi ,
\nn
\\
&&
\hat{H}'_{B^{\mu}\ph}
\;\,=\;
ie\Phi^{*}\na_{\mu}
-ie\left(
\na_{\mu}\Phi^{*}
\right)
+2e^{2}A_{\mu}\Phi^{*},
\nn
\\
&&
\hat{H}'_{\ph\, B_{\nu}}
\;\,=\;
-ie\Phi^{*}\na^{\nu}
-2ie\left(
\na^{\nu}\Phi^{*}
\right)
+2e^{2}A^{\nu}\Phi^{*},
\nn
\\
&&
\hat{H}'_{B^{\mu}\ph^{*}}
\,=\;
-ie\Phi\na_{\mu}
+ie\left(
\na_{\mu}\Phi
\right)
+2e^{2}A_{\mu}\Phi,
\nn
\\
&&
\hat{H}'_{\ph^{*}\ph^{*}}
\,=\;
-2\la\Phi\Phi,
\nn
\\
&&
\hat{H}'_{\ph\,\ph}
\;\;\;=\;
-2\la\Phi^{*}\Phi^{*},
\nn
\\
&&
\hat{H}'_{\te\, \te}
\;\;\;\;\;\!\!=\;
\cx-M_{v}^{2}.
\nn
\eeq
The matrix operator (\ref{eq:4.7}) can be rewritten as follows
\beq
\hat{H}'
&=&
\begin{pmatrix}
    -1 & 0 & 0 & 0\\
    0 & -1 & 0 & 0\\
    0 & 0 & +1 & 0\\
    0 & 0 & 0 & +1
\end{pmatrix}
\hat{H}\,,
\eeq
where the new operator $\hat{H}$ has the the canonical form
\beq
\hat{H} \,=\, \hat{1} \cx+2\hat{h}^{\al}\na_{\al}+\hat{\Pi}\,,
\eeq
with $\,\hat{1} = \diag \big(1,\,1,\,1,\, \de_{\mu}^{\nu}\big)$,
\beq
\hat{h}^{\al}
\,=\,
\begin{pmatrix}
    -ieA^{\al} & 0 & 0 & -\frac{1}{2}ieg^{\al\nu}\Phi\\
    0 & ieA^{\al} & 0 & \frac{1}{2}ieg^{\al\nu}\Phi^{*}\\
    0 & 0 & 0 & 0\\
    \frac{1}{2}ie\de_{\mu}^{\al}\Phi^{*} &
- \frac{1}{2}ie\de_{\mu}^{\al}\Phi & 0 & 0
\end{pmatrix}
\eeq
and
\beq
\hat{\Pi} =
\left(\begin{array}{cccc}
        \begin{array}{c}
            m^{2}+4\la\Phi^{*}\Phi-\xi R\\
            -ie\left(\na_{\mu}A^{\mu}\right)
            -e^{2}A^{2}_{\mu}
        \end{array}
        &  2\la\Phi\Phi
        &  0  &
        -2ie \left(\mathcal{D}^\nu\Phi\right)
\\
        2\la\Phi^{*}\Phi^{*}
        &
        \begin{array}{c}
            m^{2}+4\la\Phi^{*}\Phi-\xi R\\
            +ie\left(\na_{\mu}A^{\mu}\right)
            -e^{2}A^{2}_{\mu}
        \end{array}
        & 0 &
2ie \left(\mathcal{D}^\nu \Phi\right)^*
\\
        0 & 0 & -M_{v}^{2} & 0\\
        -ie\left(\na_{\mu}\Phi^{*}\right)
        +2e^{2}A_{\mu}\Phi^{*} & +ie\left(\na_{\mu}\Phi\right)
        +2e^{2}A_{\mu}\Phi
        & 0 &
        \begin{array}{c}
            -R_{\mu}{}_{.}^{\nu}+\de_{\mu}^{\nu}M_{v}^{2}\\
            +2\de_{\mu}^{\nu}e^{2}\Phi^{*}\Phi
        \end{array}
\end{array}\right) .
\nn
\\
\label{Pi}
\eeq

The one-loop contribution to the effective action is given by
\beq
\bar{\Ga}^{(1)}
=
\dfrac{i}{2}\textrm{Tr}\ln\hat{H}
- i\textrm{Tr}\ln\hat{H}_{\textrm{gh}},
\label{eq:4.10}
\eeq
where $\hat{H}_{\textrm{gh}}$ the operator in the ghost action
\beq
\hat{H}_{\textrm{gh}}=\cx-M_{v}^{2}.
\label{Hghost}
\eeq

The divergent part of the one-loop effective action is given
by the expression \cite{DeWitt,bavi85}
\beq
\bar{\Ga}_{\textrm{div}}^{(1)}
&=&
- \dfrac{\mu^{n-4}}{\ep}\int \! d^{n}x\sqrt{-g}\,
\,\textrm{tr}\,
\bigg\{
\dfrac{\hat{1}}{120}C^{2}
    -\dfrac{\hat{1}}{360}E_{4}
\nn
\\
&&
+ \,\,\, \dfrac{1}{2}\hat{P}^{2}
+ \dfrac{1}{12}\hat{S}_{\al\be}^{2}
+ \dfrac{1}{6}\,\cx\hat{P}
    +\dfrac{\hat{1}}{180}\cx R
\bigg\}
    \!,\,\qquad
\label{eq:4.11}
\eeq
where $C^{2}=C_{\mu\nu\al\be}^{2}$ is the square of Weyl
tensor, $E_{4}$ is the integrand of the Gauss-Bonnet topological
term,\footnote{The relations are
$R^{2}_{\mu\nu\al\be}=2C^{2}-E_{4}+\dfrac{1}{3}R^2$
and $R^2_{\mu\nu} = \frac12C^2 - \frac12E_4 + \frac13R^2$.}
 $\ep=\left(4\pi\right)^{2}\left(n-4\right)$ is the parameter
of dimensional regularization and $\mu$ is the renormalization
parameter. In the expression (\ref{eq:4.11}), the definitions are
\beq
&&
\hat{P} = \hat{\Pi}+ \dfrac{\hat{1}}{6}\,R
- \na_\al \hat{h}  - \hat{h}_\al \hat{h}^\al ,
\label{eq:4.12}
\\
&&
\hat{S}_{\al\be}
= \hat{\mathcal{R}}_{\al\be}
+ \na_\be \hat{h}_\al - \na_\al \hat{h}_\be
+ \hat{h}_\be \hat{h}_\al - \hat{h}_\al \hat{h}_\be,
    \label{eq:4.13}
\eeq
where the commutator of geometric covariant derivatives is
$\hat{\mathcal{R}}_{\al\be} =
\diag \left(0,\,0,\,0, \,- R^\mu_{\,\,\,\nu\al\be}\right)$.

The contribution of the ghost action has the standard form \cite{BuGui}
\beq
\bar{\Ga}_{\textrm{div, gh}}^{(1)}
=
-\dfrac{\mu^{n-4}}{\ep}\int d^{n}x\sqrt{-g}
\left\{ \dfrac{1}{120}C^{2}
-\dfrac{1}{360}E_{4}
+\dfrac{1}{30}\cx R
+\dfrac{1}{72}R^{2}
-\dfrac{1}{6}M_{v}^{2}R
+\dfrac{1}{2}M_{v}^{4}
\right\} .
\label{vacu}
\eeq
Using the general relation (\ref{eq:4.11}), after some algebra (the
intermediate formulas can be found in Appendix A), the divergent
part of the one-loop effective action is found in the form
\beq
&&
\bar{\Ga}_{\textrm{div}}^{(1)}
\,\,=\,\,
    -\,\dfrac{\mu^{n-4}}{\ep}
    \int d^{n}x\sqrt{-g}\,
    \bigg\{ \dfrac{1}{8}C^{2}
    -\dfrac{13}{72}E_{4}
+ \Big[\Big (\xi-\frac16\Big)^2 + \frac{1}{72}\Big] R^2
- \dfrac{1}{3}\xi\cx R
\qquad
\quad
\nn
\\
&&
\qquad\quad
    - \,\,
\Big[\dfrac{1}{6}M^{2}_{v}+2m^{2}\Big(\xi-\dfrac{1}{6}\Big)\Big]R
    + \Big[2\left(e^2-4\la\right)\Big(\xi-\dfrac{1}{6}\Big)
    - \dfrac{1}{3}e^2\Big] R \Phi^*\Phi
\nn
\\
&&
\qquad\quad
+ \,\,
\left(20\la^2-4e^2\la+4e^4\right)\left(\Phi^{*}\Phi\right)^{2}
- 2 \left[\left(e^2-4\la\right)m^2 - 3e^2M^{2}_{v}\right]
\Phi^*\Phi
    \nn
    \\
&&
\qquad\quad
+ \,\,\dfrac{2}{3}\,
\left(e^{2}+2\la\right)\cx\left(\Phi^{*}\Phi\right)
    -4e^{2}\left(D_{\mu}\Phi^{*}\right)\left(D^{\mu}\Phi\right)
    -\dfrac{e^{2}}{6}F^{2}_{\mu\nu}
    +\dfrac{3}{2}M^{4}_{v}
    +m^{4}\bigg\}.
\label{Gammadiv}
\eeq
We note that the divergences form the three groups of
terms. First of all, there are terms which reproduce the ones in
the classical action (\ref{eq:2.1}). Furthermore, there are total
derivatives terms and, finally, the vacuum terms depending only
of the external metric. This form of divergences confirms that the
violation of gauge invariance is caused only by the mass of the
Abelian vector field, i.e., the theory has only soft symmetry
breaking and this feature holds at the quantum level. We can conclude
that the renormalizability property is exactly like in the spinor
massive electrodynamics, as explored in \cite{Boulware_70}. The
general considerations of this work can be mapped to the
scalar-massive vector theory, hence the same structure of
renormalization is expected to hold in higher loop orders
up to the new non-minimal terms in vector sector.

It is worth to compare the theory under discussion not only
with the non-Abelian theory \cite{Boulware_70}, but also with
the Abelian theory of axial vector field \cite{guhesh} representing
antisymmetric torsion. In this case, the gauge symmetry is violated
by the spinor mass and, as a result, the axial vector mass is a
necessary condition of renormalizability, including in the effective
framework \cite{Beltors}. As a consequence, the longitudinal mode
of the axial vector propagates starting from the second loop
corrections and there is a conflict between renormalizability and
unitarity. Nothing of this sort occurs in the present case.

Last, but not least,  one has to take special care about the possible
gauge dependence. The reason is that, in the St\"{u}ckelberg trick
- based approach, only in the special gauge the invariant theory
reduces to the original one (\ref{eq:2.1}). In the pure massive
vector field model, even in a curved spacetime, the issue is
trivial \cite{BuGui}. However, in the present case, the story is
more complicated. What we know is that (see, e.g.,
\cite{tmf,Lavrov-Curved,OUP} for formal proofs):
\textit{i)} The divergent part of effective action is a covariant
local functional.
\textit{ii)} The power counting arguments hold on in the
theories with soft symmetry breaking \cite{Boulware_70,guhesh}.
\textit{iii)} The difference between the one-loop divergences
estimated in different gauges are proportional to the classical
equations of motion \cite{aref}. Taking together, these arguments
mean one can expect the gauge-dependent covariant local additions
to (\ref{Gammadiv}) being proportional to Eqs.~(\ref{eqPhiA}),
i.e., there may be additional terms of the form
\beq
&&
\De \bar{\Ga}_{\textrm{div}}^{(1)}
\,=\,   \dfrac{\mu^{n-4}}{\ep} \int d^n x
\sqrt{-g}\,
\Big\{ \dfrac{f_\Phi}{2} \big(
\mathcal{E}\Phi^{*} + \mathcal{E}^*\Phi\big)
+ f_\th \,\mathcal{E}_\th\, \th
+ f_A \,\mathcal{E}^\nu A_\nu\Big\},
\label{DelGa}
\eeq
where $f_\Phi$, $f_\th$ and $f_A$ are arbitrary (gauge-fixing
dependent) real functions. These parameters are completely
analogous to the constants $\al_1$, $\,\al_2$, and $\al_3$
in Eq.~(\ref{TracePhiA}).
The inspection of Eq.~(\ref{DelGa}) shows that it cannot produce
dramatic changes in the divergences, e.g., generate a longitudinal
mode of the vector field or the $(A_\mu A^\mu)^2$-type term, as
it occurs in the theory of axial vector \cite{guhesh}. This means,
there are good chances that the theory (\ref{eq:2.1}) may be
renormalizable beyond one-loop order if the proper ultraviolet
completion terms are introduced.

Replacing the equations of motion (\ref{eqPhiA}) in (\ref{DelGa}),
after some algebra we arrive at the difference between the one-loop
divergences corresponding to two  choices of the gauge fixing in
the symmetric (St\"{u}ckelberg) phase
\beq
&&
\De\bar{\Ga}_{\textrm{div}}^{(1)}
\,\,=\,\,
\bar{\Ga}_{\textrm{div}}^{(1)}\left(\chi\right)
- \bar{\Ga}_{\textrm{div}}^{(1)}\left(\chi_{0}\right)
\,\,= \,\,
\dfrac{\mu^{n-4}}{\ep}\int d^{n}x\sqrt{-g}
\,\,\bigg\{
f_\Phi
\big[\left(D^{\mu}\Phi\right)^{*}\left(D_{\mu}\Phi\right)
\nn
\\
&&
\qquad
\qquad
- \,\,\big(m^{2} - \xi R\big)\Phi^{*}\Phi
- 2\la\left(\Phi^{*}\Phi\right)^{2}\big]
\,+ \,f_{\th}\big[\left(\na_{\mu}\th\right)^{2}
    + M_{v}\th\,\na_{\mu}A^{\mu}\big]
\nn
\\
&&
\qquad
\qquad
\quad
+ \,\,\,
f_{A}\Big[\dfrac{1}{2}F_{\mu\nu}^{2}
    + M_{v}^{2}\Big(A_\nu^2 - \dfrac{1}{M_v}
A_{\nu}\na^{\nu}\th\Big)    - A_{\nu}J^{\nu}\Big]\bigg\}.
\qquad
\label{DelGa2}
\eeq
In this expression, $\chi$ correspond to an arbitrary gauge fixing
while $\chi_{0}$ represents a specific gauge condition of our choice.
The one-loop divergence for gauge fixing arbitrariness is given by sum
of the equations (\ref{Gammadiv}) and (\ref{DelGa2}). Let us note that
the requirement of gauge invariance imposes the condition
$f_\th=f_A=0$, while $f_\Phi$ is unconstrained.

\section{One-loop renormalization group equations}
\label{sec5}

Consider renormalization in the theory
(\ref{eq:2.1}). Since the calculations we performed for a different
theory (\ref{eq:4.2}) that emerges after the St\"{u}ckelberg trick,
the renormalization relations are subjects of ambiguity which was
parameterized in (\ref{DelGa2}).

The renormalized classical action is defined as ${S}_{R}= S+ \De S$,
where $\De S$ is the counterterm required to cancel the divergences.
We can simply set \ $\De S = - \,\bar{\Ga}^{(1)}_{\textrm{div}}$.
For the scalar and vector fields, the renormalization relations are
\beq
\Phi_{0} \,=\, \mu^{\frac{n-4}{2}}
\Big(1-\dfrac{4e^{2}+f_\Phi}{2\ep}\Big)\Phi
\qquad
\mbox{and}
\qquad
A^0_\al \,=\, \mu^{\frac{n-4}{2}}\Big(1+\dfrac{e^2}{3\ep}\Big)A_\al\,,
\label{renPhiA}
\eeq
which are explicitly ambiguous.  For the two masses, we meet
\beq
m_0^2
\,=\,   m^2     + \dfrac{2}{\ep}
\big[ m^2\left(3e^2-4\la\right)-3e^2M_{v}^{2}\big]
\eeq
and
\beq
M_{0,v}^2
\,=\,
M_{v}^{2} \Big(1-\dfrac{2e^2}{3\ep}\Big)\,.
\eeq
The relations for the coupling constants have the form
\beq
&&
e_0\,=\,\mu^{\frac{4-n}{2}}\left(1-\dfrac{e^{2}}{3\ep}\right)e\,,
\label{Ren_e}
\nn
\\
&&
\la_0
\,=\,   \mu^{4-n}
\Big[\la - \dfrac{1}{\ep} \left(20\la^2-12e^2\la+4e^4\right) \Big]\,,
\label{Ren_la}
\eeq
and for the nonminimal parameter we meet
\beq
\xi_0 \,=\, \xi + \dfrac{2}{\ep}
\Big[3e^2\left(\xi-\dfrac{1}{9}\right)
- 4\la\left(\xi-\dfrac{1}{6}\right) \Big].
\label{Ren_xi}
\eeq
It is remarkable that the contribution of massive vector in this
relation is \textit{not} proportional to $\xi - 1/6$, regardless there
is no mass dependence in this formula. The reason is that, before
we take a strictly massless limit, there is an extra degree of
freedom (equivalent to the St\"{u}ckelberg field), which is
non-conformal. The situation is analogous to the discontinuity
described for the vacuum sector of the massive vector field
theory in curved spacetime \cite{BuGui}.

One can find the beta and gamma functions using relations
\beq
\be_{P}\,=\,\lim_{n\rightarrow4}\, \mu\, \dfrac{dP}{d\mu}
\qquad
\textrm{and}
\qquad
\ga_{\Phi}H \,=\, \lim_{n\rightarrow4}\,\mu\,\dfrac{dH}{d\mu},
\eeq
where \ $P=\left( m,M_v,\xi,\la,e\right) $ are renormalized parameters
and \ $H=\left(\Phi^{*},\Phi,A_{\mu}\right)$ \ renormalized fields.
Using the renormalization relations, we get
\beq
&&
\be_{e}=\dfrac{e^{3}}{3\left(4\pi\right)^{2}},
\nn
\\
&&
\be_{\la}=\dfrac{1}{\left(4\pi\right)^{2}}
    \left(20\la^{2}-12e^{2}\la+4e^{4}\right),
\nn
\\
&&
\be_{\xi}=\dfrac{2}{\left(4\pi\right)^{2}}
    \Big[4\la\Big(\xi-\dfrac{1}{6}\Big)
    - 3e^{2}\Big(\xi-\dfrac{1}{9}\Big)\Big],
\nn
\\
&&
\be_{m^{2}}=-\dfrac{2}{\left(4\pi\right)^{2}}
    \left[3e^{2}\left(m^{2}-M_{v}^{2}\right)
    -4m^{2}\la\right],
\nn
\\
&&
    \be_{M_{v}^{2}}=\dfrac{2e^2}{3\left(4\pi\right)^2}M^{2}_{v}.
\label{betas}
\eeq
The gamma-functions have the form
\beq
\ga_{\Phi} \,=\,
\dfrac{2e^2}{ \left(4\pi\right)^{2}}
    \quad
    \textrm{and}
    \quad
\ga_{A_{\mu}}=-\dfrac{e^{2}}{3\left(4\pi\right)^{2}}.
\label{gamas}
\eeq
We note that $\ga_{A_\mu}$ and $\be_e$ are exactly those of the
usual QED. As we explained in Sec.~\ref{sec3}, this feature is
supposed to hold only at the one-loop order. On the other hand,
$\ga_{\Phi}$ can be a subject of ambiguity produced by the
St\"{u}ckelberg procedure in the higher-loop orders.

The effective potential for action (\ref{eq:2.1}) can be obtained
using the renormalization group (RG) technique \cite{ColeWein}
adapted to curved spacetime \cite{BuchOd-84} (see also \cite{book}).
The renormalization group equation for the effective potential
has the form
\beq
\Big\{
\mu\dfrac{\pa}{\pa\mu}
+ \be_{P}\dfrac{\pa}{\pa P}
+ \ga_\Phi \Big( \Phi \dfrac{\pa}{\pa \Phi}
+ \Phi^* \dfrac{\pa}{\pa \Phi^*} \Big)
\Big\}\,
 V_{\textrm{eff}}\left(g_{\al\be},\Phi^{*},\Phi,P,\mu\right)
\,=\,0,
\label{eq:EP}
\eeq
where $\mu$ is the renormalization parameter. The previous equation
allows the effective potential to be rewritten in terms of beta and gamma
functions. In our case, it is essential to account for the effect of the
coupling between the scalar field and the gauge field, as it results
in additional corrections to the potential. The procedure is analogous
to that described in \cite{book}, hence we just formulate the result
for the effective potential,
\beq
V_{\textrm{eff}}
&=&
V \,+\, \dfrac{1}{2}\big(\be_m  + 2m^2 \ga_\Phi \big) |\Phi|^2
\Big[ \ln\Big( \dfrac{|\Phi|^2}{\mu^2} \Big) - C_1 \Big]
\nn
\\
&&
+\,\, \dfrac{1}{2}\big(\be_\xi  + 2\xi \ga_\Phi \big) R |\Phi|^2
\Big[ \ln\Big( \dfrac{|\Phi|^2}{\mu^2} \Big) - C_2 \Big]
\nn
\\
&&
+\,\, \dfrac{1}{2}\big(\be_\la  + 4\la \ga_\Phi \big)
|\Phi|^4 \Big[\ln\Big(\dfrac{|\Phi|^2}{\mu^2}\Big) - C_3 \Big].
\label{VeffRG}
\eeq
where $|\Phi|^2=\Phi^* \Phi$, and the constants
$C_1$, $C_2$ and $C_3$ depend on the renormalization
conditions. E.g., in the massless case and the conditions used in
\cite{ColeWein,BuchOd-84,book}, the values are $C_1=0$,
$C_2=-3$ and $C_3= - 25/6$.

\section{Conclusions}
\label{sec6}

We have shown that the renormalizable  curved-space theory of
massive vector field coupled to a scalar can be based on the
minimal action (\ref{eq:2.1}), without inclusion
of nonminimal terms (\ref{eq:2.1extra}) in the vector field sector,
including those proportional to the Ricci tensor. In this respect, the
massive vector field is crucially different from the scalar field,
where the nonminimal interaction to Ricci scalar is the necessary
condition for renormalizability. The difference is that the vector
nonminimal terms are protected by the gauge symmetry, even
regardless this symmetry is softly broken in the original formulation
of the theory. The statement about renormalizability was
confirmed by the direct one-loop calculation.

The evaluation of one-loop divergences is an important part of
a QFT model, as it provides information on the renormalization
structure of the theory. We reported about the derivation of
one-loop divergences in
the massive charged scalar theory coupled to the massive vector
theory. This model proved renormalizable if the vector field is
Abelian and if it is not an axial vector. The non-Abelian vector
theory or axial vector are not renormalizable, as it was discussed
in \cite{Boulware_70} and proved by direct calculations in
\cite{guhesh}.

The beta functions (\ref{betas}) and the effective potential
(\ref{VeffRG}) were derived in the symmetric formulation
(\ref{eq:4.2}). The general arguments concerning classification
of parameters in curved spacetime \cite{OUP} imply that the
expressions (\ref{betas}) do not change under an arbitrary
choice of the gauge fixing, hence these expressions can be
applied to the original theory (\ref{eq:2.1}).

\section{Acknowledgments}
Authors are grateful to Jos\'e Abdalla~Helayel-Neto for useful
discussions. I.L.B. is thankful to D.I. Kazakov for the discussion
of renormalizabity of massive spinor electrodynamics in flat space
in the book \cite{BogShir}. P.R.B.R.V is grateful the Department
of Theoretical Physics at Zaragoza University for kind hospitality
and to Coordena\c{c}\~ao de Aperfei\c{c}oamento de Pessoal de N\'ivel
Superior - CAPES  (Brazil)
for supporting his Ph.D. project. G.Y.O. is thankful to
Funda\c{c}\~{a}o de Amparo \`a Pesquisa do Estado do Esp\'{i}rito
Santo - FAPES (Brazil) for supporting his Ph.D. project. The work
of I.Sh. is partially supported by Conselho Nacional de
Desenvolvimento Científico e Tecnológico - CNPq  (Brazil) under
the grant 305122/2023-1.

\section*{Appendix A. Intermediate expressions for one-loop divergences}

Let us collect the intermediate formulas for the derivation of
divergences in Sec.~\ref{sec3}.
According to the equations (\ref{eq:4.12}) and (\ref{eq:4.13}),
we obtain
\beq
\label{A1}
\hat{P} \,=\,    \left(P\right)_{\mu}^{\nu} \,=\,
\begin{pmatrix} \label{eq:4.7-1}
\hat{P}_{\ph^{*}\ph} & \hat{P}_{\ph^{*}\ph^{*}}
& \hat{P}_{\ph^{*}\te} & \hat{P}_{\ph^{*}B_{\nu}}
\\
\hat{P}_{\ph\,\ph} & \hat{P}_{\ph\,\ph^{*}} & \hat{P}_{\ph\,\te}
& \hat{P}_{\ph\, B_{\nu}}
\\
\hat{P}_{\te\,\ph} & \hat{P}_{\te\,\ph^{*}}  & \hat{P}_{\te\,\te}
& \hat{P}_{\te\, B_{\nu}}
\\
\hat{P}_{B^{\mu}\ph} & \hat{P}_{B^{\mu}\ph^{*}}
& \hat{P}_{B^{\mu}\te}  & \hat{P}_{B^{\mu}B_{\nu}},
\end{pmatrix}
\eeq
with the  following non-zero elements:
\beq
\label{pmunu}
&&
\hat{P}_{\ph\,\ph^{*}}
\;=\;\,
\hat{P}_{\ph^{*}\ph}
\;=\;
m^{2}   +\left(\frac{1}{6}-\xi\right)R
- \left(\frac{1}{4}\de_{\mu}^{\nu}e^{2}
- 4\la\right)\Phi^{*}\Phi,
\nn
\\
&&
\hat{P}_{\te\,\te}
\;\;\,=\; -\left(M^{2}_{v}-\frac{1}{6}R\right),
\quad
\hat{P}_{B^{\mu}B_{\nu}}
\,=\,
- R_{\mu}{}^{\nu}
+ \de_{\mu}^{\nu}\left(M_{v}^{2}
+ \frac{1}{6}R+\frac{3}{2}e^{2}\Phi^{*}\Phi\right),
\nn
\\
&&
\hat{P}_{\ph^{*}\ph^{*}}
\!\!\,\;=\;
\frac{1}{4}\left(\de_{\mu}^{\nu}e^{2}
+ 8\la\right)\Phi\Phi,
\quad\;\;
\hat{P}_{\ph\,\ph}
\;=\;
\frac{1}{4}\left(\de_{\mu}^{\nu}e^{2}
+ 8\la\right)\Phi^{*}\Phi^{*} ,
\nn
\\
&&
\hat{P}_{\ph^{*}B_{\nu}}
=\;
- \frac{3}{2}ie
\left(D^{\nu}\Phi\right),
\qquad\quad\,\,
\hat{P}_{B^{\mu}\ph}
\,=\,
- \frac{3}{2}ie
    \left(D_{\mu} \Phi\right)^{*},
    \nn
    \\
    &&
    \hat{P}_{\ph\, B_{\nu}}
    \;=\;
    \frac{3}{2}ie\left(D^{\nu}
    \Phi\right)^{*},
    \qquad\quad\,\,
    \hat{P}_{B^{\mu}\ph^{*}}
    \,=\,
    \frac{3}{2}ie
    \left(D_{\mu}
    \Phi\right).
    \nn
    \\
\eeq
Similarly, matrix $\hat{S}^{\al\be}$ is given by
\beq
\label{A2}
\hat{S}^{\al\be}
\,=\,
\left(S^{\al\be}\right)^{\nu}{}_{\mu}
\,=\,
\begin{pmatrix}
\hat{S}_{\ph^{*}\ph} & \hat{S}_{\ph^{*}\ph^{*}}
& \hat{S}_{\ph^{*}\te} & \hat{S}_{\ph^{*}B_{\nu}}
\\
\hat{S}_{\ph\,\ph} & \hat{S}_{\ph\,\ph^{*}} & \hat{S}_{\ph\,\te}
& \hat{S}_{\ph\, B_{\nu}}
\\
\hat{S}_{\te\,\ph} & \hat{S}_{\te\,\ph^{*}}  & \hat{S}_{\te\,\te}
& \hat{S}_{\te\, B_{\nu}}
\\
\hat{S}_{B^{\mu}\ph} & \hat{S}_{B^{\mu}\ph^{*}}
& \hat{S}_{B^{\mu}\te}  & \hat{S}_{B^{\mu}B_{\nu}}
\end{pmatrix} ,
\eeq
where the non-zero elements are
\beq
\label{A3}
&&
\hat{S}_{\ph^{*}\ph}
\;\,=\,\,
\frac{1}{4}e^2
\left(\de^{\al}_{\mu}g^{\be\nu}
    -\de^{\be}_{\mu}g^{\al\nu}
    \right)\Phi^{*}\Phi
    +ie\left(\na^{\al}A^{\be}
    -\na^{\be}A^{\al}\right),
\nn
\\
&&
    \hat{S}_{\ph\,\ph^{*}}
    \,\,=\,\,
        \frac{1}{4}e^2
    \left(\de^{\al}_{\mu}g^{\be\nu}
    -\de^{\be}_{\mu}g^{\al\nu}
    \right)\Phi^{*}\Phi
    -ie\left(\na^{\al}A^{\be}
    -\na^{\be}A^{\al}\right) ,
\nn
\\
&&
    \hat{S}_{B^{\mu}B_{\nu}}
    \;\!\!=\;
    R_{\mu}{}^{\nu\al\be}+
    \frac{1}{2}e^2\left(
    \de^{\be}_{\mu}g^{\al\nu}
    -\de^{\al}_{\mu}g^{\be\nu}
    \right)\Phi^{*}\Phi,
\nn
\\
&&
    \hat{S}_{\ph^{*}B_{\nu}}
    =\,\,
    \frac{1}{2}ie\left[g^{\be\nu}
    \left(D^{\al}\Phi\right)
    -g^{\al\nu}
    \left(D^{\be}\Phi\right)
     \right],
\nn
\\
&&
    \hat{S}_{B^{\mu}\ph}
    \,\,=\,
    -\frac{1}{2}ie\left[
    \de_{\mu}^{\be}\left(D^{\al}\Phi^{*}\right)
    -\de_{\mu}^{\al}\left(D^{\be}\Phi^{*}\right)
    \right],
\nn
\\
&&
    \hat{S}_{\ph\, B_{\nu}}
    \,\,=\,
    -\frac{1}{2}ie\left[g^{\be\nu}\left(D^{\al}\Phi^{*}\right)
    -g^{\al\nu}\left(D^{\be}\Phi^{*}\right)
    \right] ,
\nn
\eeq
\beq
&&
    \hat{S}_{B^{\mu}\ph^{*}}
    =\,
    \frac{1}{2}ie\left[
    \de_{\mu}^{\be}\left(D^{\al}\Phi\right)
    -\de_{\mu}^{\al}\left(D^{\be}\Phi\right)
    \right],
\nn
\\
&&
    \hat{S}_{\ph^{*}\ph^{*}}
    \,=\,
    \frac{1}{4}e^2\left(
    \de^{\be}_{\mu}g^{\al\nu}
    -\de^{\al}_{\mu}g^{\be\nu}
    \right)\Phi\Phi,
\nn
\\
&&
    \hat{S}_{\ph\,\ph\,}
    \;\;\,=\,
    \frac{1}{4}e^2\left(
    \de^{\be}_{\mu}g^{\al\nu}
    -\de^{\al}_{\mu}g^{\be\nu}
    \right)\Phi^*\Phi^*.
\nn
\eeq

Using these operators, the particular traces are
\beq
\label{BoxP}
\dfrac{1}{6}\,\textrm{tr}\,\cx\hat{P}
\,=\,
\dfrac{2}{3}\left(e^{2}+2\la\right)\cx\left(\Phi^{*}\Phi\right)
+\dfrac{1}{3}\Big(\dfrac{1}{12}-\xi\Big)\cx R,
\eeq
\beq
\label{P^2}
&&
\dfrac{1}{2}\,\textrm{tr}\,\hat{P}^2
\,=\,
\dfrac{1}{2}R^{2}_{\mu\nu}
    - \Big[\dfrac{5}{72}
    -\Big(\xi-\dfrac{1}{3}\Big)\xi\Big] R^2
    + \Big[\dfrac{M_v^2}{2} + 2 m^2 \Big(\xi-\dfrac{1}{6}\Big)\Big] R
+ \dfrac{5}{2}M_{v}^{4}
    + m^4
\nn
\\
&&
    \qquad\quad
    + \Big[2\left(e^{2}-4\la\right)
    \Big(\xi-\dfrac{1}{6}\Big)
    -\dfrac{1}{2}e^{2}  \Big]
    R\left(\Phi^{*}\Phi\right)
    +\left(20\la^{2}-4e^{2}\la
    +5e^{4}\right)\left(\Phi^{*}\Phi\right)^{2}
\nn
\\
&&
    \qquad\quad
    -2\Big{[}m^{2}\left(e^{2}-4\la\right)
    -3e^{2}M_{v}^{2}\Big{]}
    \left(\Phi^{*}\Phi\right)
    -\dfrac{9}{2}e^{2}
    \left(D^{\mu}\Phi\right)^{*}\left(D_{\mu}\Phi\right)
\eeq
and
\beq
\label{S^{2}}
\dfrac{1}{12}\,\textrm{tr}\,\hat{S}{}^{\,2}_{\al\be}
\,=\, - \dfrac{1}{12}R^{2}_{\mu\nu\al\be}
+ \dfrac{e^2}{6} R\left(\Phi^{*}\Phi\right)
- e^4\left(\Phi^{*}\Phi\right)^2
+ \dfrac{e^2}{2}\left(D^{\mu}\Phi\right)^{*}
\left(D_{\mu}\Phi\right)
- \dfrac{e^2}{6}F_{\mu\nu}^{2}.\,\;
\eeq

\section*{Appendix B. Using the auxiliary operator approach}

Let us consider an alternative approach for deriving one-loop
effective action, which was successfully applied to the pure theory
of massive vector field \cite{bavi85} and proved equivalent to
the St\"{u}ckelberg procedure - based approach. Here we shall
use similar method in the theory of massive vector coupled to
the scalar field.

First we have to determine the bilinear form of the action
(\ref{eq:2.1}) without restoring the gauge symmetry. Using the
background field method, we decompose the fields into classical
and quantum counterparts as
\beq
\Phi^{*}\rightarrow\Phi^{*}=\Phi^{*}+\ph^{*},
\quad
\Phi\rightarrow\Phi=\Phi+\ph,
\quad
A_{\mu}\rightarrow A_{\mu}=A_{\mu}+B_{\mu}
\label{bfm-Proca}
\eeq
and write the bilinear in quantum fields part of the action
in the form
\beq
S^{(2)}
&=&
\dfrac{1}{2}\int d^{4}x\sqrt{-g}\,
\begin{pmatrix}
    \ph^{*} & \ph & B^{\mu}
\end{pmatrix}
\hat{H}
\begin{pmatrix}
\ph\\
\ph^{*}\\
B_{\nu}
\end{pmatrix},
\\
\label{heis}
\mbox{where}
&&
\hat{H}
\,=\,
\begin{pmatrix}
\hat{H}_{\ph^{*}\ph} & \hat{H}_{\ph^{*}\ph^{*}}
& \hat{H}_{\ph^{*}B_{\nu}}
\\
\hat{H}_{\ph\,\ph} & \hat{H}_{\ph\,\ph^{*}}
& \hat{H}_{\ph\, B_{\nu}}
\\
\hat{H}_{B^{\mu}\ph} & \hat{H}_{B^{\mu}\ph^{*}}
& \hat{H}_{B^{\mu}B_{\nu}}
\end{pmatrix}
\eeq
and the elements of the matrix are
\beq
\label{A4}
    &&
    \hat{H}_{\ph^{*}\ph\,}
    \,\,=\;
    -\cx-m^{2}
    -4\la\Phi^{*}\Phi+\xi R
    +2ieA_{\mu}\na^{\mu}
    +ie\left(\na_{\mu}A^{\mu}\right)
    +e^{2}A^{2}_{\mu},
    \nn
    \\
    &&
    \hat{H}_{\ph\,\ph^{*}\,}
    \;=\;
    -\cx-m^{2}
    -4\la\Phi^{*}\Phi+\xi R
    -2ieA_{\mu}\na^{\mu}
    -ie\left(\na_{\mu}A^{\mu}\right)
    +e^{2}A^{2}_{\mu},
     \nn
    \\
    &&
    \hat{H}_{B^{\mu}B_{\nu}}
    \!=\;
    \de_{\mu}^{\nu}\cx
    -\na_{\mu}\na^{\nu} -R_{\mu}{}_{.}^{\nu} -\de_{\mu}^{\nu}M_{v}^{2} +2\de_{\mu}^{\nu}e^{2}\Phi^{*}\Phi ,
    \nn
    \\
    &&
    \hat{H}_{\ph^{*}B_{\nu}\,}
    =\;
    ie\Phi\na^{\nu}
    +2ie\left(\na^{\nu}\Phi\right)
     +2e^{2}A^{\nu}\Phi ,
     \nn
    \\
    &&
    \hat{H}_{B^{\mu}\,\ph\,}
    \,=\;
    ie\Phi^{*}\na_{\mu}
    -ie\left(\na_{\mu}\Phi^{*}\right)
    +2e^{2}A_{\mu}\Phi^{*},
     \nn
    \\
    &&
    \hat{H}_{\ph\,B_{\nu}}
    \,\,=\;
    -ie\Phi^{*}\na^{\nu}
    -2ie\left(\na^{\nu}\Phi^{*}\right) +2e^{2}A^{\nu}\Phi^{*} ,
    \nn
    \\
    &&
    \hat{H}_{B^{\mu}\ph^{*}\,\,}
    \!=\; -ie\Phi\na_{\mu}
    +ie\left(\na_{\mu}\Phi\right)
     +2e^{2}A_{\mu}\Phi ,
     \nn
     \\
        &&
     \hat{H}_{\ph^{*}\ph^{*}}
     \,=\;
     -2\la\Phi\Phi ,
\qquad 
     \hat{H}_{\ph\,\ph\,\,\,}
     \;\,=\;
     -2\la\Phi^{*}\Phi^{*}.
\eeq
The next step is to introduce the auxiliary operator
\beq
\label{aux}
\hat{K}
\,=\,
{K}{}_{\nu}^{\al}
\,=\,
\begin{pmatrix}
    -1 & 0 & 0 \\
    0 & -1 & 0 \\
    0 & 0 &
    - \na_{\nu}\na^{\al}
    + \de_{\nu}^{\,\al} M_v^2
\end{pmatrix}.
\eeq
The operator in the corner 
of this matrix
is the auxiliary operator introduced in \cite{bavi85}. However, in
the present case, the form of the product is much more complicated,
\beq
\label{A5}
\hat{H}^{*} \,=\, \hat{H}\hat{K} \,=\,
\begin{pmatrix}
\hat{H}^{*}_{\ph^{*}\ph} & \hat{H}^{*}_{\ph^{*}\ph^{*}}
& \hat{H}^{*}_{\ph^{*}B_{\al}}
\\
\hat{H}^{*}_{\ph\,\ph} & \hat{H}^{*}_{\ph\,\ph^{*}}
& \hat{H}^{*}_{\ph\, B_{\al}}
\\
\hat{H}^{*}_{B^{\mu}\ph} & \hat{H}^{*}_{B^{\mu}\ph^{*}}
& \hat{H}^{*}_{B^{\mu}B_{\al}}
\end{pmatrix} ,
\eeq
with the elements 
\beq 
\label{A6}
&&
\hat{H}^{*}_{\ph^{*}\ph\,}
\,\,=\;
\cx
+m^{2}
+4\la\Phi^{*}\Phi
-\xi R
-2ieA_{\nu}\na^{\nu}
-ie\left(\na_{\nu}A^{\nu}\right)
-e^{2}A^{2}_{\nu},
\nn
\\
&&
\hat{H}^{*}_{\ph\,\ph^{*}\,}
\;=\;
 \cx
 +m^{2}
 +4\la\Phi^{*}\Phi
 -\xi R
+2ieA_{\nu}\na^{\nu}
+ie\left(\na_{\nu}A^{\nu}\right)
-e^{2}A^{2}_{\nu},
\nn
\\
&&
\hat{H}^{*}_{B^{\mu}B_{\al}}
\!=\;
M_{v}^{2}
\left(
\de_{\mu}^{\al}
\cx
-R_{\mu}{}_{.}^{\al}
-\de_{\mu}^{\al}M_{v}^{2}
\right)
-2e^{2}\left(\Phi^{*}\Phi\right)
\left(
\na_{\mu}\na^{\al}
-\de_{\mu}^{\al}M^{2}_{v}
\right),
\nn
\\
&&
\hat{H}^{*}_{\ph^{*}B_{\al}\,}
=\;
-ie\Phi
\left(
\cx
\,- \,M_{v}^{2}
\right) \na^{\al}
-2ie\left(
D^{\nu}\Phi
\right)
\left(
\na_{\nu}\na^{\al}
- \de^{\al}_{\nu}M_{v}^{2}
\right),
\nn
\\
&&
\hat{H}^{*}_{B^{\mu}\,\ph\,}
\,\,=\;
-ie\Phi^{*}\na_{\mu}
 +ie\left(\na_{\mu}\Phi^{*}\right)
 -2e^{2}A_{\mu}\Phi^{*},
 \nn
\\
&&
\hat{H}^{*}_{\ph\, B_{\al}\,}
\,\,=\;
ie\Phi^{*}\left(\cx
-M_{v}^{2}\right)\na^{\al}
+2ie\left(
D^{\nu}\Phi\right)^{*}
\left(
\na_{\nu}\na^{\al}
-\de_{\nu}^{\al}M_{v}^{2}
\right),
\nn
\\
&&
\hat{H}^{*}_{B^{\mu}\ph^{*}}
\!\;\!\;=\;
ie\Phi\na_{\mu}
-ie\left(\na_{\mu}\Phi\right)
-2e^{2}A_{\mu}\Phi ,
\nn
\\
&&
\hat{H}^{*}_{\ph^{*}\ph^{*}}
\,=\;
2\la\Phi\Phi,
\nn
\\
&&
\hat{H}^{*}_{\ph\,\ph\,\,\,}
\;\,=\;
2\la\Phi^{*}\Phi^{*}.
\nn
\eeq

The form of the operator $\hat{H}^*$ in Eq.~(\ref{A5}) is quite
unusual because some of the non-diagonal elements
of the matrix have operators which are third-order differential
operators. The derivation of $\Tr \log \hat{H}^*$ in this case is a
nontrivial problem, which would be difficult to deal with even
using the generalized Schwinger-DeWitt technique of \cite{bavi85}.
Since we have another, much simpler, approach described in
Sec.~\ref{sec3}, we do not describe the elaboration of this
functional trace. This Appendix was included just to illustrate that
the two methods may be not equivalent in the more sophisticated
models with soft symmetry breaking.


\end{document}